\documentclass[useAMS,usenatbib]{mnras}
\usepackage{graphicx}
\usepackage{color}
\usepackage{amssymb}
\usepackage{epsfig}

\usepackage{comment}
\usepackage{amsmath}	
\usepackage[normalem]{ulem}

\title[On the M87 jet structure near the central engine]
{On the M87 jet structure near the central engine}

\author[V.~S.~Beskin et al.]
{\parbox{\textwidth} {V.~S.~Beskin$^{1,2}$,
T.~I.~Khalilov$^{2,1}$,
E.~E.~Nokhrina$^{1,2}$,
I.~N.~Pashchenko$^{1}$
and E.~V.~Kravchenko$^{1,2}$ } 
\\
\\
\\
$^{1}$P.N.Lebedev Physical Institute, Leninsky prospekt 53, Moscow 119991, Russia \\
$^{2}$Moscow Institute of Physics and Technology, Institutsky per.~9, Dolgoprudny 141700, Russia \\
}

\begin{document}

\date{Accepted. Received; in original form}

\pagerange{\pageref{firstpage}--\pageref{lastpage}} \pubyear{2018}

\maketitle

\label{firstpage}

\begin{abstract}
At present, there is no doubt that relativistic jets observed in active 
galactic nuclei pass from highly magnetized to  weakly magnetized stage, 
which is observed as a break in the dependence on their width $d_{\rm jet}(z)$ 
on the distance $z$ to the central engine. In this paper, we discuss 
the possibility of observing another break, which should be located at 
shorter distances. The position of this break can be associated 
with the region of formation of the dense central core near the jet 
axis which was predicted both analytically and numerically more than
a decade ago, but has not yet received sufficient attention. In this case,
the observed width should be determined by the dense core, and not by the 
total transverse size of the jet. The calculations carried out in this
paper, which took into account both the transverse electromagnetic structure
of the jet and the change in the spectrum of emitting particles along its
axis, indeed showed such behaviour. We also found the evidence of the predicted 
break in the jet expansion profile using stacked 15 GHz VLBA image of M87
radio jet and constrain the light cylinder radius.
\end{abstract}



\section{Introduction}
\label{Sect1}

A real breakthrough in the study of the internal structure of relativistic jets 
emanating from active galactic nuclei (AGN), achieved over the past decade thanks 
to the high angular resolution obtained using very long baseline interferometry 
(VLBI) observations~\citep{Kovalev1, MOJAVE_V, MOJAVE_XII,MOJAVE_XIII, Mertens}  
allows us to investigate directly their internal structure. In particular, 
it give us the direct information about the dependence of  the jet width 
$d_{\rm jet}(z)$ on the distance $z$ to the central engine.  
It finally became possible to compare the results of observations with the 
predictions of a theory, the basic elements of which have long been 
constructed~\citep{B76, L76, Ardavan, BZ-77, BBR84, Camenzind86,
HN89, Camenzind90, Tomimatsu, PP92, BP93, Heyvaert96}.

We recall that the nature of relativistic jets from AGN is believed to be 
associated with highly magnetized magneto-hydrodynamical (MHD) outflows 
generated by fastly rotating supermassive black holes~\citep{Krolik, 
Camenzind07, MHD, Meier}. Within this approach, the key role is played 
by the poloidal magnetic field generated in the accretion disc as within 
this model both plasma outflow and the energy flux propagate along magnetic 
field lines from the central engine to active regions. As was shown 
by~\citet{BZ-77}, a strongly magnetized flow (in which the main role in 
energy transfer is played by the electromagnetic field, i.e. the Poynting 
vector flux) is really capable of taking the energy away from a rotating
black hole and transmitting it to infinity. As a result, thanks to numerous 
works devoted to both analytical and numerical analysis of MHD 
equations~\citep{Chiueh91, Appl92, Eichler93, Bogovalov95, Ustyugova95, 
Beskin97, LHAN99, BM00, VK03, McKinney06, BN06, Komissarov07, BN09, Romanova, 
Lyu09, MHD, Tchekhovskoy_11, Porth_etal11, McKinney12, PC15, BCKN-17},
a consistent model was constructed, which is now generally accepted.

Let us note from the very beginning the following key points characterizing 
the MHD model of relativistic jets; the assumptions of the model we use 
will be described in detail in Section 2. First, it should be stressed the role 
of the ambient pressure $P_{\rm ext}$ which actually determines the transverse
size of the jet $d(z)$ and, hence, their inner structure~\citep{Beskin97, Lyu09}.
In particular, it is the pressure dependence 
$P_{\rm ext}(z)$ on the distance $z$  that is to determine the dependence of 
transverse size of the jet on the distance from the central engine. 

Second, as was found both analytycally~\citep{BN06, Lyu09} and 
numerically~\citep{McKinney06, Komissarov07, Tchekhovskoy_11, Porth_etal11},
for a sufficiently high external pressure $P_{\rm ext}$ (i.e. not so
far from the central engine), the longitudinal poloidal magnetic field 
within the jet is to be homogeneous. However, as the external pressure 
decreases, a denser core begins to form in the very centre of the jet, 
while the poloidal magnetic field $B_{\rm p}$ and the number density 
$n_{\rm e}$ of the outflowing plasma begin to decrease significantly 
with distance $\varpi$ from the jet axis. But the flow still 
remains highly magnetized. Recently this result was comfirmed 
by~\citet{zero} for relativistic jets with zero velocity along 
the axis.

Third, as in the magnetically dominated flow the energy of particles 
increases with the distance $\varpi$ from the jet axis, at large 
enough distances $z$ corresponding to lower external pressures 
(where the jet width $d_{\rm jet}(z)$ also becomes large enough), 
inevitably, saturation should occur, when almost all the 
electromagnetic energy flux (which, as was already stressed, is 
dominated near the black hole) will be transferred to the energy  
of outflowing plasma (see Sect.~\ref{Sect2} for more detail).

Currently, there are direct indications that saturation occurs at the 
distances of several parsecs ($10^5$--$10^6$ of gravitational radii 
$R_{\rm g} = GM/c^2$) from the central engine~\citep{NKP20}. Such 
a transition from highly magnetized to weakly magnetized regime was 
predicted to be observed as a break in the dependence on the width of 
the jet $d_{\rm jet}(z)$ on the distance $z$~\citep{BCKN-17, NGBNAH19}. This 
results in the observed parabolic streamline in the upstream region,
and the jet shape becomes conical downstream a few $10^5~R_{\rm g}$. 
This structural transition of a jet shape was first discovered in 
M87~\citep{2012ApJ...745L..28A} and the similar transition 
was lately found in a number nearby 
sources~\citep{2016ApJ...833..288T, 2018ApJ...860..141H, 2018ApJ...854..148N, 
2018Galax...6...15A, Kovalev2020, 2020AJ....159...14N}.

Remember that for M87 such a nature of the break has also been 
discussed by~\citet{Mertens}. But they talked about a feature 
located at a distance about 10 mas, i.e., hundreds of times smaller. 
Therefore, it is of interest to discuss the question of whether 
the feature at the very base of the jet in the M87 is not connected
with  the formation of the central core, i.e. with the transition
from the jet with homogeneous poloidal magnetic field to the jet 
having a dense central core. An indication of such a break can be 
found in the work of~\citet{2016ApJ...817..131H}. However, this 
issue has not yet been discussed in detail. 

Thus, in this paper, we investigate the internal structure of relativistic 
jets at small distances $z$ from the central engine. Having decided on
the model of the internal structure of electromagnetic fields (which, within 
the framework of the MHD approach, allows us to determine the plasma number 
density as well), we then find the observed width of the jet, the radiation 
of which is associated with the synchrotron radiation of outflowing particles. 
Our goal is to find out whether, in addition to external break in the dependence 
of the width of the relativistic jet $d_{\rm jet}(z)$ on the distance $z$, at which the 
transition from magnetically dominated to particle dominated flow occurs, 
there is also an internal break in which the central core near the jet 
axis begins to form. 

Let us recall that the source of radiating leptons are currently unknown. 
Since the seminal paper by~\citet{BZ-77} it has been believed that they can be
formed as a result of the collision of thermal gamma rays emitted by the highly 
heated internal regions of the accretion disk. Numerical simulations for pair
production from the thermal seed radiation of plasma in a vicinity of a black 
hole event horizon by~\citet{Monika} provide the number of electron-positron 
pairs enough to explain the jet particle number density implied by a core-shift 
method (see e.g.~\citealt{NBKZ15} and references therein) further downstream. 
It is assumed \citep[see e.g.][]{BlandfordKonigl79,1981ApJ...243..700K}
that the jet emission in radio is due to synchrotron self-absorbed radiation. 
In optically thin regime far from a radio core the observed power-law 
spectrum~\citep{2014AJ....147..143H,2014MNRAS.437.3396K} can be explained by the
power-law energy distribution of emitting particles. The standard approach usually 
separates the MHD jet modelling and the mechanisms of emitting  plasma acceleration 
and formation of an expected power-law  spectrum~\citep[see e.g.][]{Kirk1994, SironiSpitkovskyArons2013,
Ostrowski1990, Ostrowski1998}. This is due to the complexity of the full problem 
of merging MHD with physical kinetics~\citep[particle-in-cell method, 
see][ for a review]{Kagan2015} in one scheme. \citet{Sironi2016} superimposed the development 
of blobs of highly relativistic plasma, forming due to reconnection, on a jet-like
structure. But modelling the spectral flux maps to explain the observational data
is achieved at the moment by imposing the assumed non-thermal plasma distribution 
onto the jet structure~\citep{Porth_etal11, 2012MNRAS.423..756P, 2013MNRAS.429.1189P, 
2019MNRAS.488..939P, 2019A&A...629A...4F, Ogihara19, KramerMacDonald2021, FNP23}. 
In this paper, we follow the latter approach.

The paper is organized as follows. In Section~\ref{Sect2} the basic 
relations of the MHD theory are formulated, allowing a detailed 
description of the internal structure of relativistic jets.  Based on 
this theory, in Section~\ref{sec:jet_transverse_dimension} we determine
the transverse size of relativistic jets near their base, which allows
us to find the position of the internal break. Finally, in 
Section~\ref{sec:observations} we employ the observational data on the 
break and discuss the imposed constraints.

\section{Internal structure of relativistic jets}
\label{Sect2}

First of all, let us dwell in detail on our model of the internal 
structure of relativistic jets. It is based on the results obtained 
by~\citet{BN06}, according to which, for strongly collimated jets, one 
can consider their internal structure as a sequence of cylindrical flows
depending on a jet radius $\varpi$ only. For such cylindrical configurations, 
the Grad-Shafranov (GS) equations describing internal structure of ideal MHD 
flows (see~\citealt{HN89, Camenzind90, PP92, BP93, Heyvaert96} for more 
detail) can be easily integrated~\citep{Beskin97, BM00, Lyu09} as it reduces 
to two ordinary differential equations of the first order. As a result, 
by specifying four integrals of motion at the base of a magnetically
dominated flow (see below), it is possible to obtain transverse profiles 
of electromagnetic fields (and, within the framework of this approach, 
hydrodynamic speed and particle number density as well) at an arbitrary
distance from the central engine.

\begin{figure} 
\begin{center}
\includegraphics[width=0.95\columnwidth]{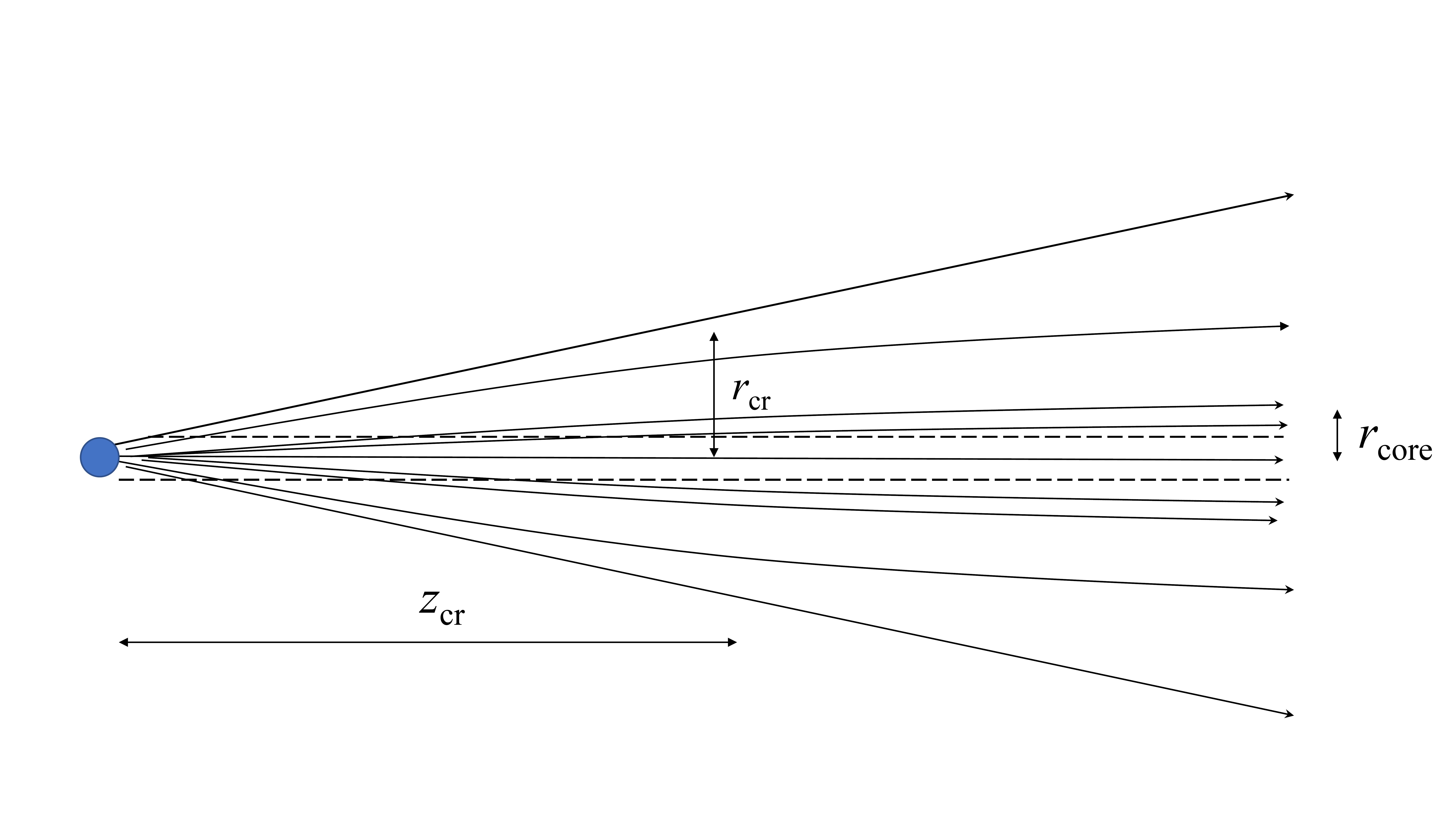}
  \end{center}
  \caption{The structure of the magnetic field in the conical model under consideration. 
  At a distances $z > z_{\rm cr}$ from the central engine,  when the transverse radius of 
  the jet reaches the scale $\sim r_{\rm cr}$, in a conical flow (in which the plasma 
  density and the magnetic field weakly depend on the distance from the axis), a denser 
  central core begins to form. The light cylinder $\varpi = c/\Omega_{0}$ is shown by a 
  dashed line.
  }
\label{Figur2}
\end{figure}

A detailed discussion of the model used can be found in our previous 
works~\citep{BN09, NBKZ15, BCKN-17, PaperI}. Here we recall only two 
significant circumstances. First, as already noted, one of the important
features of the model is that at sufficiently large distances from the 
central engine, a denser core begins to form in the very center of the 
jet, and the poloidal magnetic field $B_{\rm p}$ and the number density 
$n_{\rm e}$ of the outflowing plasma begin to decrease significantly 
with distance $\varpi$ from the jet axis. This structure, shown in 
Figure~\ref{Figur2}, was also confirmed by numerical 
simulation~\citep{Komissarov07, Tchekhovskoy_11, Porth_etal11}.

Second, despite the cumbersome form of the GS equation,
rather simple asymptotic solutions have been formulated, which will 
be enough for us to state the main points of this paper. As was already stressed, 
the structure of the poloidal magnetic field within relativistic jets 
outflowing from active galactic nuclei substantially depends on the 
ambient pressure $P_{\rm ext}$ which, in turn, depends on the distance 
$z$ from the central engine. Therefore, in what follows, we consider 
all quantities as functions of the distance $z$. 

\begin{figure} 
\begin{center}
\includegraphics[width=0.9\columnwidth]{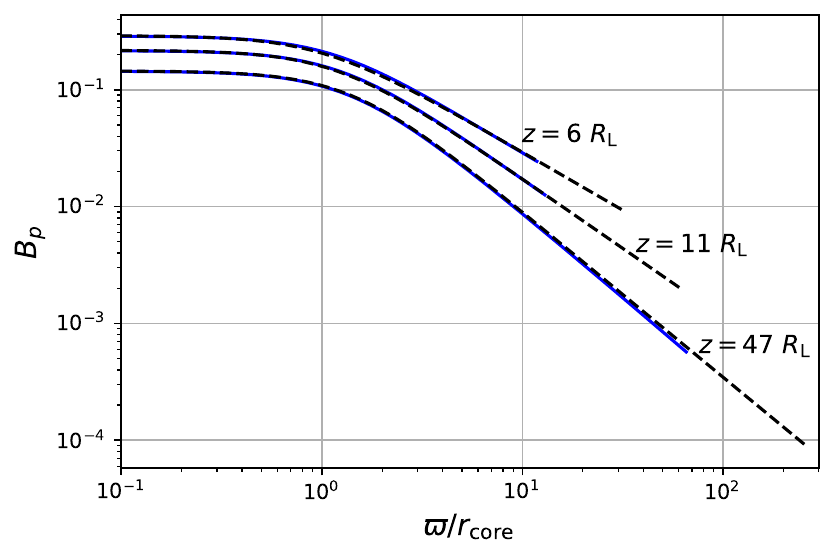}
  \end{center}
  \caption{Comparison of analytical approximation (\ref{I})--(\ref{II}) (black dashed lines) with the structure of a poloidal magnetic field (blue solid lines) obtained by solving a system of two ordinary differential equations describing internal structure of a cylindrical jet. In this figure, we terminated the solid lines of numerical solution at smaller radii so that the analytical approximation may be better seen. We plot three jet crosscuts for a conical jet with intrinsic opening angle $10^{\circ}$. 
  }
  \label{Figur1}
\end{figure}

As a result, with a good accuracy, poloidal magnetic field $B_{\rm p}$
can be represented as
\begin{eqnarray}
B_{\rm p}(\varpi) \approx  B_{\rm p}(0), \qquad \qquad \qquad \quad \quad B_{\rm p}(0) > B_{\rm cr},  
\label{I} \\
 B_{\rm p}(\varpi) \approx B_{0}(z) \left(1 + 
 \frac{\varpi^2}{r_{\rm core}^2}\right)^{-\alpha(z)/2},   B_{\rm p}(0) < B_{\rm cr}.
\label{II} 
\end{eqnarray}
Here
\begin{equation}
r_{\rm core} = \gamma_{\rm in} R_{\rm L}
\label{rcore}
\end{equation}
is the core radius where the integration constant $\gamma_{\rm in} \sim 1$ corresponds 
to the Lorentz-factor of a flow along the axis, and $R_{\rm L} = c/\Omega_{0}$ is the radius 
of the light cylinder (the constant $\Omega_{0}$ will be defined below). 

In other words, at small distances from the central engine, poloidal magnetic field 
remains actually constant (so that $B_{0}^2/8\pi \approx P_{\rm ext}$){\footnote{This 
relationship is valid in the case when the longitudinal current is closed inside the 
jet itself, which we are considering here.}}, while at large distances it decreases 
with distance from the axis $\varpi$. As was already stressed, this result has been 
repeatedly confirmed by numerical calculations~\citep{Komissarov07, Tche08, Porth_etal11}. 
Moreover, this structure is realized for zero hydrodynamical velocity along the jet axis 
($\gamma_{\rm in} = 1$) as well~\citep{zero}. As shown in Figure~\ref{Figur1}, analytical 
approximation (\ref{I})--(\ref{II}) perfectly reproduces the structure of a poloidal 
magnetic field obtained by solving a system of two ordinary differential equations 
describing internal structure of cylindrical jet (see~\citealt{BN09, Lyu09, PaperI} 
for more detail).

Wherein, the characteristic magnetic field $B_{\rm cr}$, at which the formation of the central 
core begins, can be presented as 
\begin{equation}
B_{\rm cr} = \frac{B_{\rm L}}{\gamma_{\rm in}\sigma_{\rm M}},
\label{Bcr}
\end{equation}
where
\begin{equation}
\sigma_{\rm M} = \frac{\Omega_{0}^2\Psi_{\rm tot}}{8 \pi^2 \mu \eta c^2}
\label{sigmaM}
\end{equation}
is so-called~\citet{Michel69} magnetization parameter. Here $\mu \approx m_{\rm e}c^2$
is the relativistic enthalpy, and $\eta = nu_{\rm p}/B_{\rm p}$ is the particle-to-magnetic 
flux ratio\footnote{{Unfortunately, at present there is no reliable data concerning 
the dependence of the integral of motion $\eta = \eta(\Psi)$ on the magnetic flux $\Psi$.
Therefore, in what follows we assume $\eta = $ const.}}. Due to numerous evaluations (see,
e.g.~\citealt{NBKZ15}), for most jets from active galactic nuclei the Michel magnetization 
parameter \mbox{$\sigma_{\rm M} \sim 10$--$50$.} For M87 we assume $\sigma_{\rm M} \sim 20$. 
In turn, the magnetic field $B_{\rm L}$ is determined from the relation 
\begin{equation}
\Psi_{\rm tot} = \pi R_{\rm L}^2 B_{\rm L},
\label{BL}
\end{equation}
where $\Psi_{\rm tot}$ is the total magnetic flux within the jet.
In what follows we assume $B_{\rm L} = 100$ G, which corresponds to
$R_{\rm L} = 10 R_{\rm g}$.

Further, the exponent $\alpha$ changes from 0 to 1 as the ambient 
pressure decreases to values $P_{\rm eq} = B_{\rm eq}^2/8\pi$, where 
\mbox{$B_{\rm eq} = \sigma_{\rm M}^{-2}B_{\rm L}$.} In this region,
the magnetic field on the axis $B_{0}$ weakly depends on $P_{\rm ext}$.
In all this magnetically dominated region, the asymptotic behaviour
$\gamma(x) = x = \Omega \varpi/c$ is fulfilled. 
At an even lower ambient pressure $P_{\rm ext} < B_{\rm eq}^2/8\pi$, 
the flow becomes partically dominated. In this domain $\alpha \approx 1$. 

It should be emphasized that in order to determine the explicit form of 
dependences $B_{0}(z)$ and $\alpha(z)$, it is necessary to specify either 
the dependence of ambient pressure $P_{\rm ext}(z)$ on $z$ or the shape 
of the jet boundary. However, since the ambient pressure is not known with
sufficient accuracy, below we use the second approach. It is based on the 
fact that the parameter of the problem considered by~\citet{BN06} was not 
the distance from the central engine $z$, but the transverse radius of 
a jet $r_{\rm jet}$. Therefore, by specifying the shape of the jet, one can 
also determine the dependence of the parameters on the coordinate $z$. Below 
we use the conical model (see Figure~\ref{Figur2}), which seems to be the most 
reasonable at the very base of the jet. As we see below, this assumption
does not contradict the observed parabolic shape of the jets.

Using now the results of the work of~\citet{BN06}, we obtain the following 
approximation for the functions $\alpha\left(z\right)$ and $B_0\left(z\right)$ 
defining the poloidal magnetic field (\ref{I})--(\ref{II}) in the case of a 
conical structure for $z > z_{\rm cr}$
\begin{eqnarray}
\alpha\left(z\right) & \approx &
0.05\ln\left(z/z_{\rm cr}\right), 
\label{alpha} \\
B_0\left(z\right) & \approx & B_{\rm cr} \frac{1 - \alpha\left(z\right)/2}{\left(1+z^2/z_{\rm cr}^2\right)^{1-\alpha\left(z\right)/2}-1}.
\label{B0}
\end{eqnarray}
In (\ref{B0}) $B_0\left(z\right)$ is determined from the conservation of the total magnetic flux. 

From here it can be seen that at short distances, before the formation of the 
central core, we have $B_0\left(z\right)\, \propto\, z^{-2}$. It should be noted
that the approximate function \eqref{alpha} at distances $z\gg z_{\rm cr}$ becomes
more than one, which contradicts the condition in the region when the flow of 
electromagnetic energy almost completely passes into the plasma energy flow. 
Nevertheless, within the framework of this paper, such distances will not be 
considered, and this question concerns an approximate function such that the 
inequality $0 < \alpha\left(z\right) < 1$ is satisfied.

The advantage of the approach considered here (in contrast to works in which it was necessary 
to solve a system of two ordinary differential equations) is that the expression (\ref{II}) allows 
one to write down the magnetic flux \mbox{$\Psi = 2 \pi \int B_{\rm p} \varpi\, {\rm d} \varpi$} 
explicitly
\begin{equation}
\Psi(\varpi, z) = \frac{2 \pi B_{0}(z)r_{\rm core}^2}{2-\alpha(z)} 
\left[\left(1 + \frac{\varpi^2}{r_{\rm core}^2}\right)^{1 - \alpha(z)/2} - 1\right].
\label{Psi}
\end{equation}
In turn, it allows us to find all the integrals of motion in any point, as they 
in standard GS approach~\citep{HN89, PP92} depend on magnetic flux $\Psi$  only. 

Indeed, the key role in the MHD theory of strongly magnetized jets play the integrals~\citep{MHD}
\begin{eqnarray}
E(\Psi) & = & \gamma \mu \eta c^2 + \frac{\Omega_{\rm F}I}{2\pi,} 
\label{E} \\ 
L(\Psi) & = & \varpi u_{\varphi} \mu \eta c + \frac{I}{2\pi},    
\label{L}
\end{eqnarray}
where $E(\Psi)$ (integral Bernoulli) corresponds to the energy flux and $L(\Psi)$ to the 
angular momentum flux. Here $\Omega_{\rm F}(\Psi)$ is the angular velocity which is also
constant on magnetic surfaces (Ferraro's isorotation law), and $I$ is the total current 
inside the magnetic surface. The value  $\Omega_{0}$ we introduced earlier is equal to 
$\Omega_{\rm F}(0)$ by definition. Besides, in the black hole magnetosphere both the 
electric current $I$ and the angular velocity $\Omega_{\rm F}$ are determined from the  
critical conditions on the singular surfaces. Wherein, the current $I$ entering the 
relations (\ref{E})--(\ref{L}) turns out to be close to the so-called Goldreich-Julian 
current (appropriate current density $j_{\rm GJ} = \Omega_{\rm F}B_{0}/2\pi$), and the 
angular velocity $\Omega_{\rm F}$ is close to half the angular velocity of the black hole 
rotation $\Omega_{\rm H}$.

Thus, with good accuracy, the Bernoulli integral can be written in the form
\begin{equation}
E(\Psi) \approx \Omega_{\rm F}(\Psi)L(\Psi) + 
\gamma_{\rm in}\mu\eta c^2,
\label{EPsi}
\end{equation}
and besides, in the region of strongly magnetized flow, the contribution 
of particles can be neglected. On the other side, as we know, for both conical~\citep{Michel73} 
and parabolic~\citep{B76} flows near the axis, we can set $L(\Psi) = \Omega_{0}\Psi/(4\pi^2)$.
In what follows, however, we use the integrals~\citep{NGBNAH19, PaperI}
\begin{eqnarray}
L(\Psi) & = & \frac{\Omega_{0}\Psi}{4 \pi^2}\sqrt{1 - \frac{\Psi}{\Psi_{\rm tot}}}, 
\label{Ldef} \\ 
\Omega_{\rm F}(\Psi) & = & \Omega_{0}\sqrt{1 - \frac{\Psi}{\Psi_{\rm tot}}},
\label{OmegaF}
\end{eqnarray}
which allow us to include in consideration the reverse electric current flowing within the jet.

As a result, knowing the potential $\Psi(\varpi, z)$ and explicit expressions for the integrals 
of motion $\Omega_{\rm F}(\Psi)$, $E(\Psi)$, and $L(\Psi)$, we can determine all components of 
electric and magnetic fields within the jet
\begin{eqnarray}
\bmath{B}_{\rm p} & = & \frac{\nabla \Psi \times \bmath{e}_{\varphi}}{2 \pi \varpi}, 
\label{Bp} \\ 
B_{\varphi} & = & - (1 + \varepsilon)\frac{\Omega_{\rm F}(\Psi)}{2\pi c} |\nabla \Psi|, 
\label{Bphi} \\ 
\bmath{E}  & = & -  \frac{\Omega_{\rm F}(\Psi)}{2\pi c} \nabla \Psi.
\label{Er}
\end{eqnarray}
Here, however, we add the additional factor $(1 + \varepsilon)$ into the expression
for the toroidal magnetic field, where $\varepsilon=\varepsilon(\Psi)$ is a small 
parameter, $\varepsilon \ll 1$. This increase of the  toroidal magnetic field with 
respect to exact force-free solution makes it possible to simulate the absence of
particle acceleration at large distances, when the electromagnetic energy flux has 
already been transferred to the plasma flow. Moreover, such a correction does not 
violate the basic relationship 
\mbox{$\bmath{E} + (\bmath{\Omega}_{\rm F}\times \bmath{r}/c)\times \bmath{B} = 0$.}

Indeed, using the fundamental theoretical result, according to which, outside the light 
cylinder, the hydrodynamic velocity $\bmath{V}$ becomes almost equal to the electrical drift 
velocity~\citep{Tche08, MHD}
\begin{equation}
\bmath{U}_{\rm dr} = c \frac{\bmath{E}\times \bmath{B}}{B^2},
\label{Udrdef}
\end{equation}
so the hydrodynamic Lorentz-factor of plasma can be written as $\Gamma = \Gamma_{\rm dr}$ where
\begin{eqnarray}
\Gamma_{\rm dr}(x) = \frac{1}{\sqrt{1 - U_{\rm dr}^2/c^2}} =
\left(1 - \frac{E^2}{B_{\varphi}^2 + B_{\rm p}^2}\right)^{-1/2} 
\nonumber \\
= \left[1 - \frac{\omega^2}{(1 + \varepsilon)^2\omega^2 + 1/x^2}\right]^{-1/2} \approx
\left(2\varepsilon + \frac{1}{\omega^2x^2}\right)^{-1/2}.
\end{eqnarray}
Here $x = \varpi/R_{\rm L}$ is the dimensionless distance to the jet axis,
and $\omega = \Omega_{\rm F}(x)/\Omega_{0}$. As a result, at small distances from 
the central engine we get for $\omega = 1$
\begin{equation}
\Gamma_{\rm dr} \approx x,
\label{Gx}
\end{equation}
i.e., well-known asymptotic solution for collimated magnetized jets~\citep{Beskin97, Lyu09, Tchekhovskoy_11}. 
As already emphasized, this asymptotic behaviour remains valid in the entire region $B_{\rm ext} > B_{\rm eq}$. 
On the other hand, at large distances, i.e. for $x > (2\varepsilon)^{-1/2}$, we have 
$\Gamma_{\rm dr} \approx (2\varepsilon)^{-1/2} \approx$ const. This asymptotic behaviour models 
the saturation region, when the entire energy flow is already concentrated in the hydrodynamic 
particle flow and Lorentz-factor of the flow reaches its maximal value 
$\Gamma_{\rm max} \approx (2\varepsilon)^{-1/2}$.

As a result, the function $\varepsilon(\Psi)$ can be easily found from Bernoulli 
equation (\ref{E}) which just determines the maximum Lorentz-factor on a given 
magnetic surface: 
\begin{equation}
\Gamma_{\max}(\Psi) = \frac{E(\Psi)}{\mu\eta c^2}.
\end{equation}
Using now relations (\ref{E})--(\ref{L}), we obtain
\begin{equation}
\varepsilon(\Psi) = \left(1 - \frac{1}{\Gamma_{\rm max}^2(\Psi)}\right)^{-1/2} - 1 
\approx \frac{1}{2\Gamma_{\rm max}^2(\Psi)},
\label{vare2}
\end{equation}
where now due to (\ref{Ldef})--(\ref{OmegaF})
\begin{equation}
\Gamma_{\rm max}(\Psi) = \gamma_{\rm in} + 
2 \sigma_{\rm M}\frac{\Psi}{\Psi_{\rm tot}}\left(1 - \frac{\Psi}{\Psi_{\rm tot}}\right),
\label{vare2bis}
\end{equation}
and the value of $\sigma_{\rm M}$ (\ref{sigmaM}) has already been defined above.
Below we use the velocity field obtained to deternime the Doppler factor.

\section{Observed jet transverse dimension}
\label{sec:jet_transverse_dimension}

Let us now discuss the observed transverse dimension of a jet $d_{\rm jet}(z) = 2 r_{\rm jet}(z)$ 
at an arbitrary distance $z$ from the central engine (see Figure~\ref{Figur2}).
Due to relation (\ref{Bcr}), the appearance of the central core occurs at the 
moment when the transverse radius of the jet becomes of order
\begin{equation}
r_{\rm cr} \sim (\gamma_{\rm in}\sigma_{\rm M})^{1/2}R_{\rm L}.
\label{r1}
\end{equation}
Accordingly, the distance $z_{\rm cr}$ to the central engine can be evaluated as
\begin{equation}
z_{\rm cr} \sim \Theta_{\rm jet}^{-1}(\gamma_{\rm in}\sigma_{\rm M})^{1/2}R_{\rm L}.
\label{z1con}
\end{equation}
For reference, we present here the corresponding estimates for parabolic flow as well
\begin{equation}
r_{\rm jet} \sim \left(\frac{z}{R_{\rm L}}\right)^{1/2}R_{\rm L}.
\label{rparabolic}
\end{equation}
\begin{equation}
z_{\rm cr} \sim \gamma_{\rm in}\sigma_{\rm M} \, R_{\rm L}.
\label{z1par}
\end{equation}
As for the value $r_{\rm cr}$, it does not depend on the geometry of the jet. 
This important result will be used significantly below.

It is clear that as long as the longitudinal magnetic field (and hence the 
number density) remains uniform within the jet, observed transverse
radius of the conical jet $r_{\rm jet}(z)$ increase linearly with distance $z$ as the total transverse radius
\begin{equation}
r_{\rm jet} \approx \Theta_{\rm jet} z.
\label{rconic}
\end{equation}
Here $\Theta_{\rm jet}$ is the total angular half-width of the conical jet. 
On the other hand, observed width of the jet $d_{\rm jet}(z)$, due to the decrease in 
the integrals of motion (\ref{E})-(\ref{L}) and the number density towards 
its edge at large distances $z > z_{\rm cr}$, is to be smaller than $2\Theta_{\rm jet}z$. 
Nevertheless, below we use relations (\ref{r1})-(\ref{z1con}) for the estimate of
the observed jet width for small distances up to $z = z_{\rm cr}$.

To more accurately determine the observed transverse size of the jet at any distance 
from the central engine, we use standard relations~\citep{BIP03, PIB03, LPB03, LPG}, 
allowing to construct the brightness temperature maps. It is based on the 
hypothesis of the power-law spectrum of radiating particles in the comoving 
reference frame 
\mbox{${\rm d}N = n_{\gamma}^{\prime} f(\gamma^{\prime}) 
\, {\rm d}\gamma^{\prime} \, {\rm d}^3{r}^{\prime} \, {\rm d}\Omega/4\pi$,}
when the energy spectrum $f(\gamma^{\prime})$ can be presented in the form
\begin{equation}
f(\gamma^{\prime}) = K_{\rm e} \, (\gamma^{\prime})^{-p}.
\label{ne}
\end{equation}
Here $\gamma_{0} < \gamma^{\prime} < \gamma_{\rm max}$, $n_{\gamma}^{\prime}$ 
is the number density of radiating particles in the comoving reference frame, 
$p$ is the spectrum index, ${\rm d}\Omega$ is the solid angle, and $K_{\rm e}$ 
is the normalising constant resulting from the condition $\int f(\gamma^{\prime}) {\rm d}\gamma^{\prime} = 1$.

Here two important clarifications must be made. First, below
we take into account the evolution of the spectrum of radiating particles
due to conservation of the first adiabatic invariant~\citep{inv}
\begin{equation}
I_{\perp} = \frac{(p_{\perp}^{\prime})^2}{h},
\label{Iperp}
\end{equation}
where
\begin{equation}
h = \sqrt{B^2 - E^2} 
\end{equation}
is a magnetic field in the comoving reference frame; for magnetically dominated flow 
considered here one can put $h \approx B_{\rm p}$. Since the condition 
$p_{\perp}^{\prime} \approx m_{\rm e}c \gamma^{\prime}$ 
is satisfied for ultrarelativistic particles, we finally obtain
\begin{equation}
\gamma^{\prime} = \frac{I_{\perp}^{1/2}}{m_{\rm e}c} \, h^{1/2}.
\label{gamma-h}
\end{equation}
Assuming now that $\gamma_{\rm max}^{\prime} \gg \gamma_{0}$ and $\gamma_{0} \gg 1$
we obtain for the normalising constant $K_{\rm e}$
\begin{equation}
K_{\rm e} = (p - 1) n_{\gamma}^{\prime}({\bmath r}^{\prime}) 
\, \gamma_{0}^{p - 1}({\bmath r}^{\prime}).
\label{Ke_exp}
\end{equation}

Thus, when moving away from the central engine, due to the preservation 
of the first adiabatic invariant (\ref{Iperp}), the entire spectrum of 
radiating particles will shift towards low energies without changing
its shape. Accordingly the normalising constant $K_{\rm e}$ will have 
the form
\begin{equation}
K_{\rm e} \propto n_{\gamma}^{\prime} \,  h^{(p-1)/2}.
\label{Ke}
\end{equation}
It is easy to check that expression (\ref{Ke}) remains the same if the lower
limit corresponds to non-relativistic velocities ($\gamma_{0} = 1$). This is 
due to a decrease in the number of ultrarelativistic particles.

Second, despite the fact that the conservation of the first adiabatic invariant 
formally leads to a noticeable angular anisotropy of the angular distribution 
of radiating particles~\citep{inv}, turbulence and plasma instabilities should
lead to isotropization of the distribution function. Therefore, in what follows
we do not take into account the effects of anisotropy of the angular distribution
of radiating particles.

On the other hand, below we are not limited to the approximation of an optically
thin plasma, although this approximation is also valid for sufficiently high 
frequencies. The point is that we are interested in regions close enough to the 
central engine, where the effects of synchrotron self-absorption can play a 
significant role~\citep{FNP23}.

Thus, to determine the brightness temperature, we use the standard radiative transfer equation
for the spectral radiance $I_{\nu}$
\begin{equation}
\frac{{\rm d}I_{\nu}}{{\rm d}l} = a_{\nu} - \mu_{\nu} I_{\nu},
\label{dIdt}
\end{equation}   
where
\begin{equation}
a_{\nu} = \frac{(p + 7/3)}{(p + 1)} \kappa(\nu) 
\frac{{\cal D}^{2 + (p -1)/2}}{\left(1+{\rm z}\right)^{2 + (p -1)/2}} |h\sin{\hat \chi}|^{(p + 1)/2}
\label{munu}
\end{equation}
is the synchrotron emissivity~\citep{Zhelez, LPB03}, and 
\begin{equation}
\mu_{\nu} = \left(p+10/3\right) \zeta(\nu)
\frac{{\cal D}^{\left(p+4\right)/2 -1}}{\left(1+{\rm z}\right)^{\left(p+4\right)/2 -1}} |h\sin{\hat \chi}|^{(p + 2)/2}
\label{mu}
\end{equation}
is the absorption opacity.
Here
\begin{equation}
{\cal D} = \frac{1}{\Gamma(1 - {\bmath \beta} {\bmath n})}
\end{equation}
is the Doppler-factor which we can now determine at any point, knowing the spatial 
distribution of hydrodynamic velocities ${\bmath V}(\bmath r)$ and the corresponding
Lorentz-factors $\Gamma(\bmath r)${\footnote{As was already stressed, outside 
the light cylinder the hydrodynamic velocity ${\bmath V}(\bmath r)$ is close to t
he drift velocity ${\bmath U}_{\rm dr}$ (\ref{Udrdef}) whereas the drift velocity 
${\bmath U}_{\rm dr} = (c/4\pi) {\bmath E} \times {\bmath B}$ in turn 
is determined from relations (\ref{Psi}) and (\ref{Bp})--(\ref{Er}).}, 
\begin{eqnarray}
\kappa(\nu) = \frac{\sqrt{3}}{4} \Gamma\left(\frac{3p - 1}
{12}\right)\Gamma\left(\frac{3p + 7}{12}\right)\, \frac{e^3}{m_{\rm e}c^2}\,
\nonumber \\
\times \left(\frac{3e}{2\pi m_{\rm e}c}\right)^{(p - 1)/2} \nu^{-(p - 1)/2}K_{\rm e},
\end{eqnarray} 
and
\begin{eqnarray}
\zeta(\nu) = \frac{\sqrt{3}}{4} \Gamma\left(\frac{3p+2}{12}\right) \Gamma\left(\frac{3p+10}{12}\right) 
\nonumber \\
\times \frac{e^3}{2m_{\rm e}^2 c^2} \left(\frac{3e}{2\pi m_{\rm e} c}\right)^{p/2} \nu^{-\left(p+4\right)/2} K_{\rm e}.
\end{eqnarray}
Using now relation $T_{\rm br} = I_{\nu}c^2/(2 k_{\rm B}\nu^2)$, we obtain
\begin{eqnarray}
T_{\rm br} = \frac{R(p)}{(1 + z)^{2 + (p-1)/2}}  \frac{e^3}{m_{\rm e}k_{\rm B} } \, \left(\frac{e}{m_{\rm
e}c}\right)^{(p-1)/2} \nu^{-(p+3)/2} 
\nonumber \\
\times \int_{0}^{\infty} {\cal D}^{2 + (p -1)/2} h^{(p +
1)/2}
n_{\gamma}(\bmath{r}) \, \gamma_{0}^{p - 1}(\bf{r})(\sin{\hat \chi})^{(p + 1)/2} 
\label{TTT} \\
\times \exp\left(-\int_{0}^{l}\mu_{\nu}(l^{\prime}){\rm d}l^{\prime}\right)
{\rm d}l,
\nonumber
\end{eqnarray}
where
\begin{equation}
R(p) = \frac{3^{p/2}(p - 1)(p + 7/3)}{8(2\pi)^{(p-3)/2}(p + 1)}
\Gamma\left(\frac{3p - 1}{12}\right)\Gamma\left(\frac{3p + 7}{12}\right),
\end{equation}
and the integration goes from the point of the observer towards the source.
Here, in contrast to~\citet{inv}, we have added the cosmological factor $(1 + z)$.

In what follows, it is convenient to write down the number density of high-energy particles
$n_{\gamma}(\bmath{r})$ entering in (\ref{TTT}) (which now corresponds to the laboratory 
reference frame) as
\begin{equation}
n_{\gamma} = \lambda_{\gamma} n_{\rm GJ},
\label{ngamma}
\end{equation}
where  
\begin{equation}
n_{\rm GJ} = \frac{|\bmath{\Omega} \bmath{B}_{\rm p}|}{2 \pi c e}
\label{nGJ}
\end{equation}
is the so-called~\citet{GJ} number density (i.e., the lowest particle density required
to screen the longitudinal electric field), and the constant $\lambda_{\gamma}$ 
is the multiplicity factor for radiating particles. The convenience of relation 
(\ref{ngamma}) is that for highly collimated flows (when the poloidal magnetic field 
$\bmath{B}_{\rm p}$ is actually directed along the rotation axis) it can be used at 
any distance from the central engine. Wherein, the entire dependence of number density
$n_{\gamma}(\bmath{r})$ on coordinates is contained in $\bmath{B}_{\rm p}(\bmath{r})$.

It is clear that the number density of emitting particles 
$\lambda_{\gamma}n_{\rm GJ}$ should not exceed the number density of the outflowing plasma 
$\lambda n_{\rm GJ}$, for which, as estimates show (see, e.g.~\citealt{Monika, NBKZ15}), 
$\lambda \sim 10^{11}$--$10^{13}$. Therefore, below we assume that the number density 
of emitting particles is about one percent of the density of the runaway plasma, so that 
$\lambda_{\gamma} = 10^{10}$. As a result, we finally get
\begin{eqnarray}
T_{\rm br} = \lambda_{\gamma} \frac{m_{\rm e}c^2}{k_{\rm B}} \frac{R(p)}{2
\pi}  \left(\frac{eh_{0}}{m_{\rm e}c}\right)^{(p+3)/2} \frac{\nu^{-(p+3)/2}}{(1 + z)^{2 + (p-1)/2}} 
\nonumber \\
\int
{\cal D}^{2 + (p -1)/2} \left(\frac{h}{h_{0}}\right)^{p} \, \frac{B_{0}(z)}{h_{0}} \, (\sin{\hat \chi})^{(p + 1)/2} 
\label{Tbr} \\ 
\frac{\Omega_{\rm F}(\Psi)}{\Omega_{0}} \, \left(1 + \frac{\varpi^2}{r_{\rm core}^2}\right)^{-\alpha(z)/2} \, \exp\left(-\int_{0}^{l}\mu_{\nu}(l^{\prime}){\rm d}l^{\prime}\right)
\frac{{\rm d}l}{R_{\rm L}}.
\nonumber
\end{eqnarray}
Here $h_{0}$ is the magnetic field at the light cylinder.

\begin{figure} 
\begin{center}
\includegraphics[width=0.8\columnwidth]{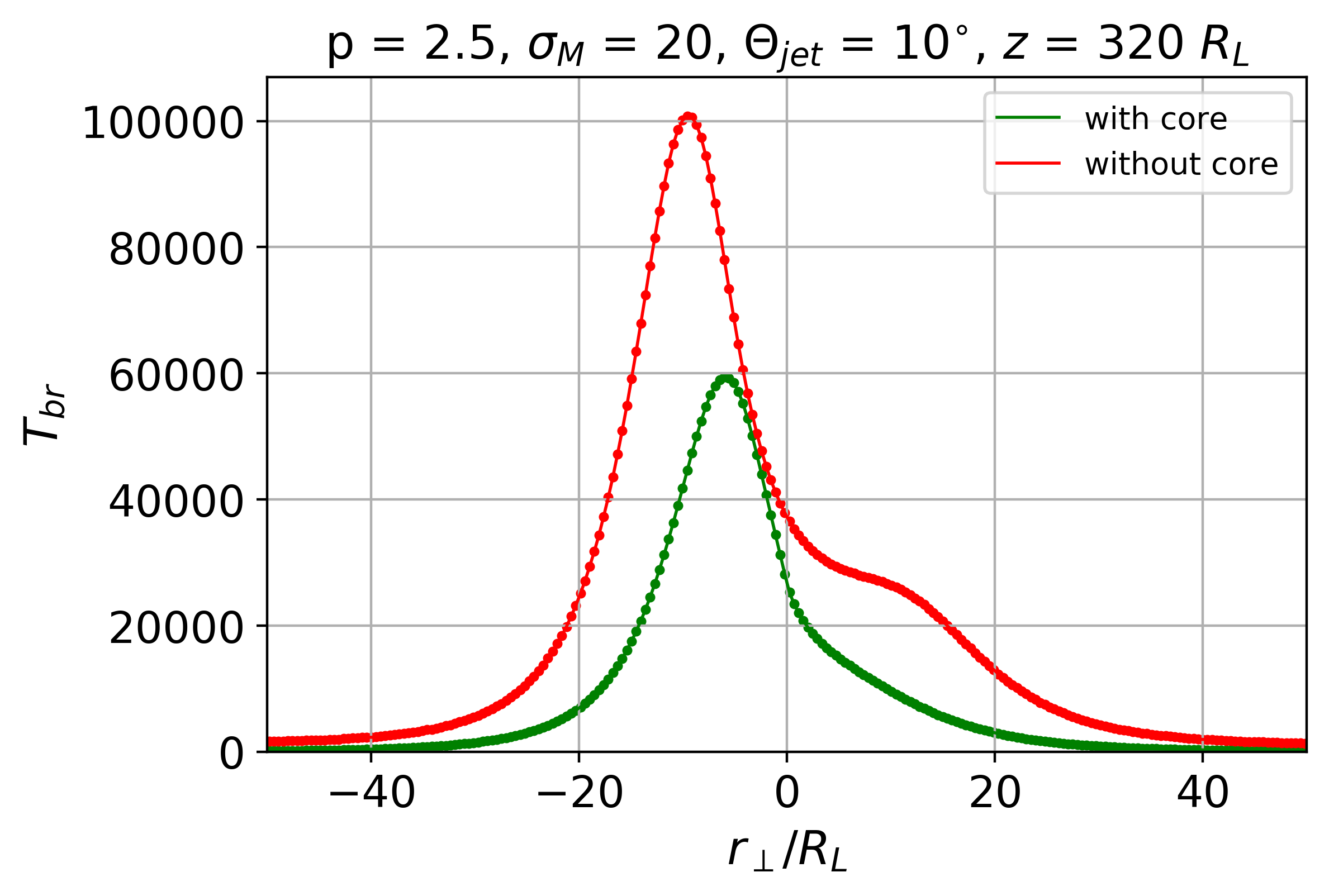}
  \end{center}
  \caption{Transverse brightness temperature profiles (spectrum index $p = 2.5$, magnetization parameter $\sigma_{\rm M} = 20$,
 and the jet angular half-width $\Theta_{\rm jet} = 10^{\circ}$) at $z = 320 \, R_{\rm L}$ from the central engine assuming the presence and the absence of a dense central core.}
\label{Figur3}
\end{figure}

\begin{figure} 
\begin{center}
\includegraphics[width=0.8\columnwidth]{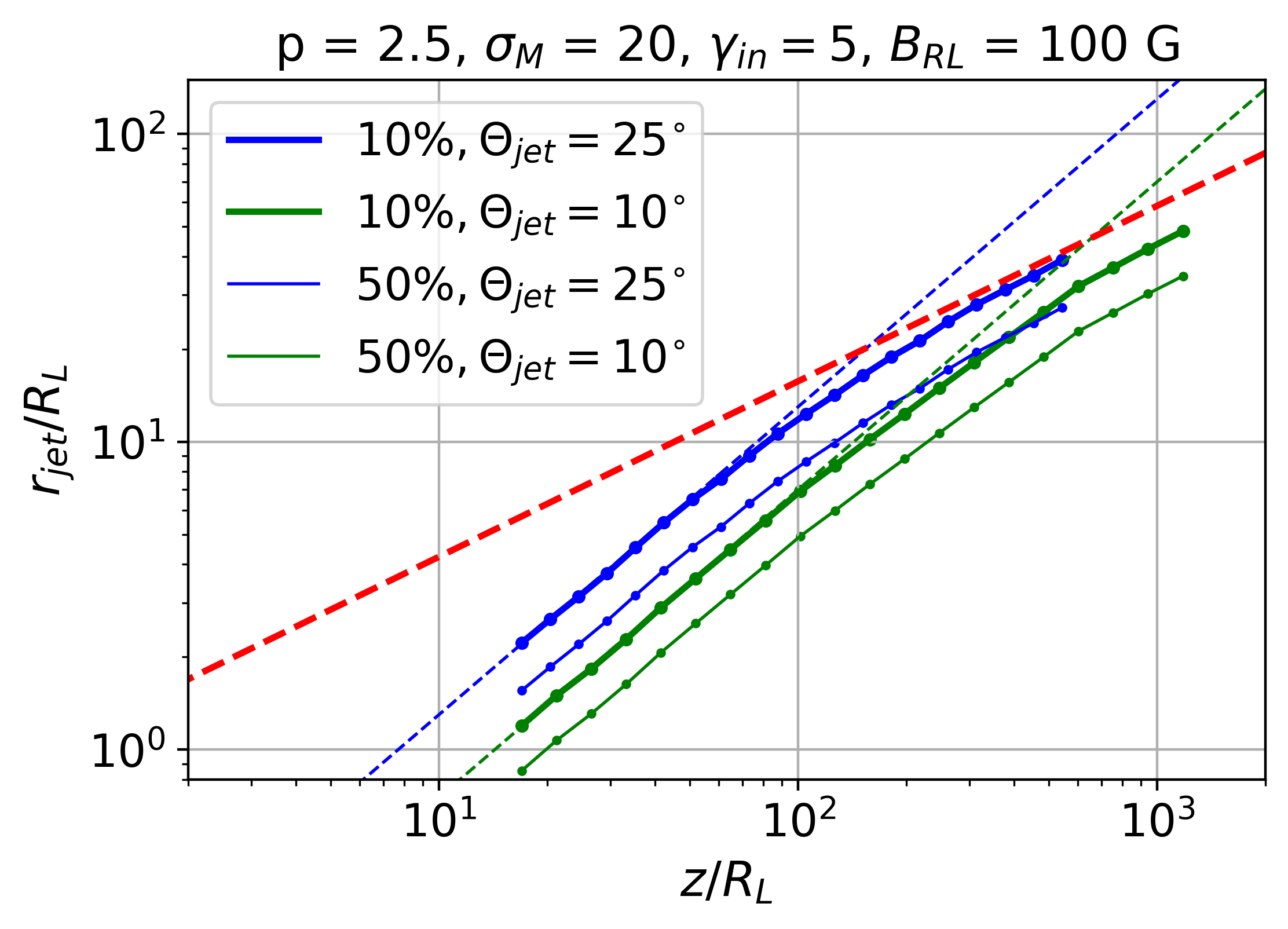}
  \end{center}
  \caption{Dependence of the observed jet width $d_{\rm jet}(z)$ at the levels of 10\%  and 50\%
   of the maximum brightness temperature for two half-widths $\Theta_{\rm jet}$ of conical jets.
   Parabolic asymptotic behaviour \mbox{$d_{\rm jet} \approx 0.06 \, z^{0.57}$ pc}~\citep{NGBNAH19} 
   shown by the bold dashed line. Thin dashed lines indicate asymptotic behaviour 
   $d_{\rm jet} \propto z$ for small $z$.}
\label{Figur4}
\end{figure}

Figure~\ref{Figur3} shows an example of predicted cross sections of 
the brightness temperature $T_{\rm br}$ for characteristic jet values 
(magnetization parameter $\sigma_{\rm M} = 20$, jet angular half-width 
$\Theta_{\rm jet} = 10^{\circ}$, spectrum index $p = 2.5$) for 
$z = 320 \, R_{\rm L}$ (which is larger than $z_{\rm cr} \approx 30 \, R_{\rm L}$) 
from the central engine assuming the presence and the absence of a dense 
central core. Everywhere in the calculations it is assumed $\nu = 15$ GHz.
As we see, the presence of a central core does result in a significant 
reduction in the observed transverse size of the jet. Recall that 
the observed asymmetry is associated with the presence of helical magnetic 
field \citep{2011MNRAS.415.2081C} and toroidal plasma velocity, due to 
which the maximum Doppler factor is achieved at a certain distance from 
the jet axis~\citep{Ogihara19,FNP23, inv}. Also note that our chosen 
value of $\lambda_{\gamma} = 10^{10}$ results in reasonable brightness 
temperature $T_{\rm br}$. 

In addition, Figure~\ref{Figur4} shows the dependence of the observed (at 
the levels of 10 and 50 percent of the maximum brightness temperature) jet 
width $d_{\rm jet}(z) = 2\Theta_{\rm j}(z)z$ for two jet half-thicknesses 
$\Theta_{\rm jet}$. As we see, at a distance of $\sim 100 R_{\rm L}$ from 
the origin, i.e. in full agreement with our evaluation (\ref{z1con}), the 
dependence $d = d(z)$ has a break. If at small distances the observed jet
width follows the conical expansion of the flow $d \propto z$  (thin dashed 
lines), then at large distances it becomes noticeably smaller. In this case, 
the width of the jet in the kink region really does not depend on $\Theta_{\rm jet}$. 
It is also interesting to note that the width of the jet approaches the 
parabolic asymptotic behaviour $d_{\rm jet} \approx 0.06 \, z^{0.57}$ pc~\citep{NGBNAH19}, 
shown by the dashed line. Thus, the formation of a central core should 
indeed lead to a significant change in the dependence of the observed jet 
width $d_{\rm jet}(z)$ on the distance $z$ to the central engine.

\section{Observations and analysis}
\label{sec:observations} 

According to Section~\ref{sec:jet_transverse_dimension}, the appearance of the core is accompanied by the break in the jet geometrical profile. Jet in the radio galaxy M87 provides the best opportunity to search for the predicted phenomenon due to its proximity. The mass of the black hole in M87 $M_\textrm{BH} \approx 6.5\times10^9 M_\odot$ and distance is 16.8 Mpc \citep{2019ApJ...875L...1E} provide the angular to spatial scale ratio of $1~\mathrm{milli arcsecond ~(mas)} \approx 0.08~\mathrm{pc} \approx 260 ~R_{\rm g}$. Thus, the expected position of the break (\ref{r1}) corresponds to its apparent position from tens of $\mu$as for large spins to few mas for moderate spin values $a\approx0.1$. 

\subsection{Searching for the break}
\label{sec:searching}

For the analysis of the internal jet structure, we considered publicly available\footnote{\url{https://www.cv.nrao.edu/MOJAVE/sourcepages/1228+126.shtml}} 15\,GHz VLBA observations obtained between 1995 July 28 and 2022 September 29 in frame of the MOJAVE program \citep{2019ApJ...874...43L}, with inclusion of data from NRAO archive. Single epoch images often poorly represent large-scale structure and stacking of the single epoch images obtained within sufficiently large time interval could reveal the whole jet cross-section \citep{2017MNRAS.468.4992P}. Jet of M87 is known to be a hard target for the CLEAN \citep{CLEAN} imaging procedure. \cite{2023MNRAS.523.1247P} showed that CLEAN produces a spurious central brightening in the intrinsically edge-brightened jets that could be suppressed by convolving CLEAN model with a larger beam. Also, M87 parsec-scale jet appearance at 15\,GHz is dominated by the helical threads of the Kelvin-Helmholtz instability \citep{2023MNRAS.526.5949N}. These introduce significant problems in the analysis of the jet structure. Thus, we convolved each single epoch CLEAN model with a common circular beam of size 1.4 mas which corresponds to typical size of the naturally weighted beam major axis. This reduces CLEAN imaging artifacts, while decreases the resolution, thus, making our results more conservative. We also filtered out 8 single epochs with a significant image noise to increase the final stacking image fidelity \citep{2023MNRAS.520.6053P}. 

To extract the collimation profile we considered the distribution of the total intensity in the image domain along the slices, taken transverse to the jet direction, which was chosen to be at position angle $PA = 17^{\circ}$ with respect to the Right Ascension axis\footnote{This value was chosen by considering the jet transverse slices from $z_{\rm obs} = 0.5$ to 30 mas.}. The slices were taken starting from 0.5\,mas from the core in steps of 0.1\,mas, and for the each cut we fitted two Gaussians to the brightness profile and taking the jet width $D$ as profile width at 10\% of the maximal fitted profile value. For the complex resolved asymmetric transverse structure of the M87 jet, where multiple components can have significant brightness difference, the usage of the half maximum level (FWHM) can be misleading by measuring the width of one component instead of the whole jet \citep{2023MNRAS.526.5949N}. As an additional check, we employed simulations with a known parabolic jet model to ensure that usage of 10\% level does not bias the geometry estimate.
The deconvolved jet cross section therefore was computed as $d_{\rm jet} = \sqrt{D^2 - \theta_{10\%}^2}$, where $\theta_{10\%}$ is the full width of the restoring beam at 10\% of the maximum.

In the deriving the jet expansion profile from the images one has to consider the dependence of the nearby pixels. This could affect the uncertainty estimation if the data points are assumed independent and identically distributed or even bias the parameters of the fit. To overcome this we considered a correlated noise model, where the noise is modelled via \textit{Gaussian Process} \citep[hereafter GP,][]{RasmussenW06}.
We employed the Rational Quadratic kernel that contains a mixture of scales.

To search the possible jet break, we employed the following relations for the jet width \citep{Kovalev2020}:
\begin{equation}
\begin{split}
d_{\rm jet}(z_{\rm obs}) = a_1(z_{\rm obs}+z_0)^{k_1},~{\rm if}~z<z_{\rm break},\\
d_{\rm jet}(z_{\rm obs}) = a_2(z_{\rm obs}+z_1)^{k_2},~{\rm if}~z>z_{\rm break},
\end{split}
\label{eq:rdbreak}
\end{equation}
Here $z_{\rm break}$ is the apparent distance of the break from the jet origin and $z_0 > 0$ corresponds to a separation of the 15\,GHz core component from the true jet origin. That is necessary in order to better fit the jet shape near the jet apex~\citep{Kovalev2020}. The coefficient $a_2$ is chosen for two power-laws to join each other at $z_{\rm break}$. We also forced the transition region to be smooth by cubic interpolation model predictions on both sides of the break at distance 0.5 mas. Both single and double slope (i.e. with a break) dependencies were fitted to the data using the \textit{Diffusive Nested Sampling} algorithm \citep{brewer2011} implemented in the \texttt{DNest4} package \citep{JSSv086i07} for sampling the posterior distribution of the model parameters. We made use the Bayesian factors for the model selection \citep{2008ConPh..49...71T}.

\begin{figure} 
\begin{center}
\includegraphics[width=1.0\columnwidth]{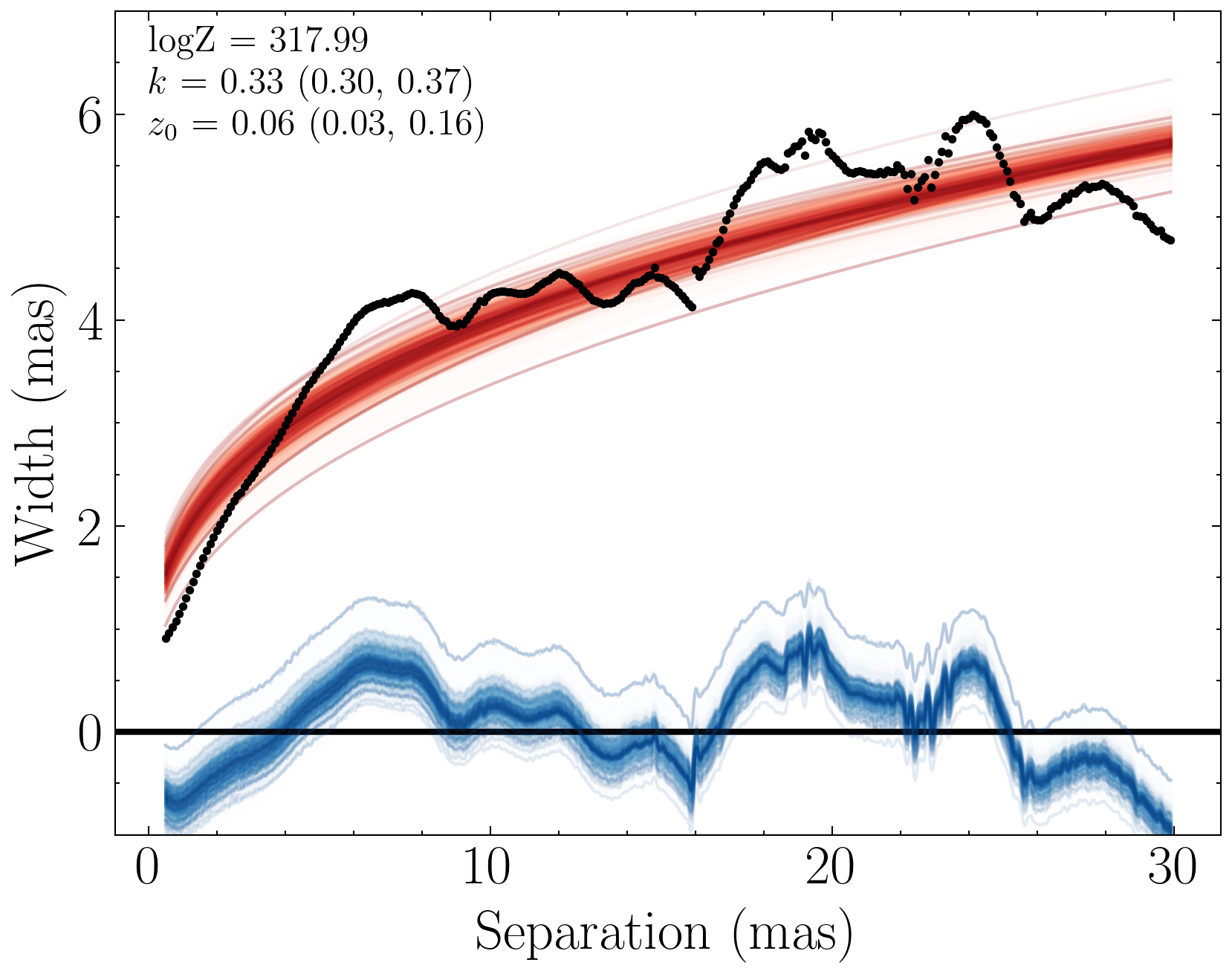}
\includegraphics[width=1.0\columnwidth]{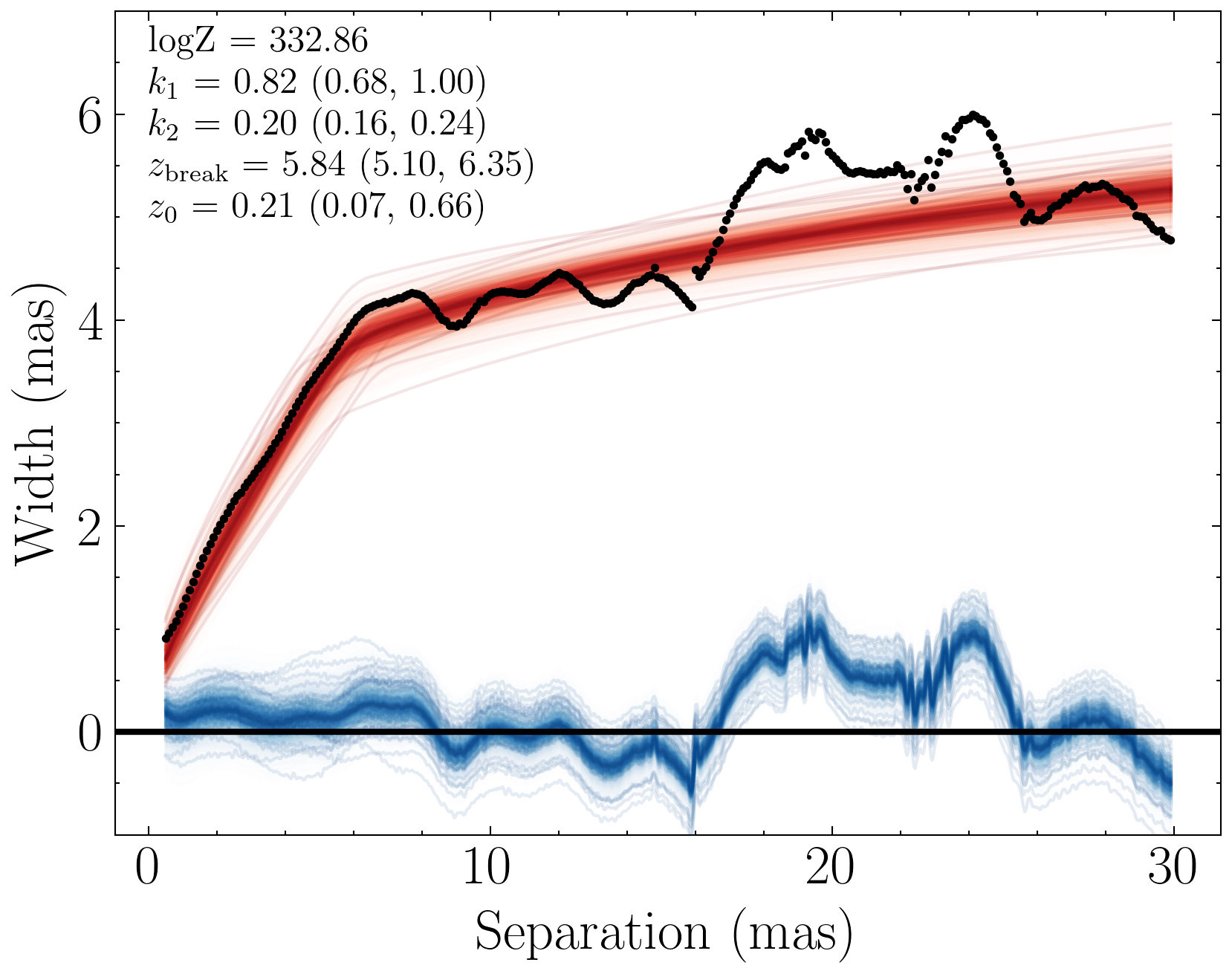}
  \end{center}
  \caption{Jet width versus apparent (projected) distance along the jet $z_{\rm obs}$ fitted by a single (\textit{Upper}) and broken (\textit{Lower}) power-law models. Black points show data points. Lines represents 500 samples from the posterior distribution. Red colour corresponds to the power-law model, while blue colour shows Gaussian Process component.}
\label{fig:gp_fit_changepoint}
\end{figure}

Accounting for the correlated noise led to an increase in the uncertainty interval width of the break position by a factor of two, and for the errors of the slope --- by four to five times. The resulting fits are presented in Figure~\ref{fig:gp_fit_changepoint} where the red colour represents the power-law model (\ref{eq:rdbreak}) and the blue colour corresponds to the GP component. The residual GP component significantly deviates from zero, especially in $z_{\rm obs} > 15$ mas region. This could be a real effect or the result of the limited $uv$-coverage. The break at $z_{\rm obs} = 5.80^{+0.65}_{-0.62}$ mas is significant with logarithm of the Bayes factor $\Delta \log{Z} = 14.9$, where $Z$ is the evidence or the marginal likelihood of the model ~\citep{2008ConPh..49...71T}. The power-law before the break is consistent with almost conical expansion, while it becomes significantly flatter after the break. Although the break seems to be a robust feature (e.g. it is clearly seen in independent high-sensitivity 15\,GHz image of~\citealt{2023MNRAS.526.5949N}, see their Figure 4), the exact values of the exponents of the expansion profile (\ref{eq:rdbreak}) should be treated with caution. \cite{2019MNRAS.488..939P} showed that at least for a relativistic jet model of \cite{BlandfordKonigl79} the jet shape measured from the stacked VLBI images could deviate systematically from the true unobserved value.

We assessed the robustness of our break detection procedure with several tests, including simulations with a known model brightness distribution. For this we created artificial multi-epoch $uv$-data set using the edge-brightned jet model without the break from \cite{2023MNRAS.523.1247P} and the same $uv$-coverages and noise as for the real data from the MOJAVE archive. After imaging the artificial data and stacking single epoch images we employed the same procedure of the break estimation as described above for the real observed data. Below we summarize our tests:

\begin{itemize}

    \item We fitted the synthetic data based on the smooth parabolic jet model with a single power-law and different number of Gaussians (single, double and triple) describing the transverse profile. The resulting fits are generally consistent with the intrinsic parabolic expansion profile $k = 0.5$ with single Gaussian fit providing the worst estimate $k^{\rm 1G} = 0.42_{-0.02}^{+0.02}$ and double and triple Gaussian fit are consistent with the true value ($k^{\rm 2G} = 0.46^{+0.03}_{-0.02}$ and $k^{\rm 3G} = 0.47^{+0.01}_{-0.03}$).

    \item 
    We found that the residual oscillations of the jet width in the synthetic stacked images of the intrinsically smooth model significantly deviate from zero at $z_{\rm obs} < 2$ mas and $z_{\rm obs} > 15$ mas. Thus, imaging artifacts could also contribute to the observed oscillations of the M87 jet width as well as helical threads of Kelvin-Helmholtz instability observed in e.g. \cite{2023MNRAS.526.5949N} or oscillations of the expanding magnetized jet \citep{Lyu09,2015ComAC...2....9K,Mertens}.

    \item 
    We also fitted the observed transverse profiles with a single and three Gaussians instead of two. The position of the break is stable against different number of Gaussians. The single Gaussian fit results in insignificantly larger break distance $z^{\rm 1G}_{\rm break} = 6.24^{+0.39}_{-0.36}$ mas than a triple Gaussian fit $z^{\rm 3G}_{\rm break} = 5.81^{+0.56}_{-0.60}$ mas.

    \item 
    The fit of the synthetic data based on the smooth jet model revealed a significant -- with even larger Bayes factor $\Delta \log{Z} = 22.4$ than for the real data -- break from the conical to the parabolic jet shape with $z^{\rm 2G}_{\rm break} = 2.20^{+0.50}_{-0.49}$ mas. Interesting that this type of imaging artefact is opposite to the systematics described in \cite{Kovalev2020} who observed that jet profiles at $z_{\rm obs} < 0.5$ mas could be artificially flattened. The fits with single and triple Gaussians also showed a break: $z_{\rm break}^{\rm 1G} = 3.26^{+0.72}_{-0.58}$ with $\Delta \log{Z} = 21.9$ and much less significant $z_{\rm break}^{\rm 3G} = 2.24^{+0.65}_{-0.77}$ mas with modest $\Delta \log{Z} = 2.6$.

    \item 
    We also checked more the influence of the convolving beam size on the break appearance and employed circular beam size 0.85 mas, that is the typical equivalent area circular beam size from the analysis of \cite{2017MNRAS.468.4992P}. The real data fit showed a significant (with $\Delta \log{Z} = 5.2$) break at $z_{\rm break}^{\rm 2G} = 4.44^{+1.77}_{-2.49}$ mas. The synthetic data revealed $z_{\rm break}^{\rm 2G} = 1.37^{+0.22}_{-0.38}$ mas with modest significance $\Delta \log{Z} = 2.7$. 
    
\end{itemize}

We conclude that the detected change of the jet shape from conical to parabolic could be affected by the imaging systematics. Thus, the constraints from the break position in the MOJAVE 15\,GHz data obtained in the following section should be treated with caution, but could be regarded as a firm limits.

\subsection{Constrains from the break position}
\label{sec:constrains}

Adopting~\cite{2019ApJ...875L...1E} estimation for M87 black hole mass \mbox{$M_\textrm{BH}=(6.5 \pm 0.9)\times10^9 M_\odot$} and the distance $(16.8 \pm 0.8)$ Mpc, we can use the observed jet width at the observed break position \autoref{r1} and its uncertainty to estimate the width of the allowed region in ($\sigma_{\rm M}$, $R_{\rm L}$) space. As emphasized above, this parameter does not depend on the jet geometry and therefore can be used for both conical and parabolic flows.

Further constrains could be obtained by considering the observed apparent motion at the position of HST-1 \citep{1999ApJ...520..621B} and assuming that it corresponds to the maximal Lorentz-factor $\Gamma_{\rm max}$ in the jet observed at the viewing angle $\Theta = 17.2\pm3.3^\circ$ \citep{Mertens}. Equation B13 from \cite{2022MNRAS.509.1899N} connects $\Gamma_{\rm max}$ and the initial magnetization $\sigma_{\rm M}$ for the considered MHD model. Using the observed jet width $d_{\rm break} = 3.8 \pm 0.3$ mas at the position of the break in the MOJAVE data estimated in \autoref{sec:searching} and assuming $\gamma_{\rm in} \approx 1$, we obtain the constraints presented in \autoref{fig:constrains_observational}. Here we also plot the constrained region that corresponds to the end of the conical expansion profile at $z_{\rm obs} \approx 0.6$ mas observed in \cite{2016ApJ...817..131H}.

\begin{figure} 
\begin{center}

\includegraphics[width=1.0\columnwidth]{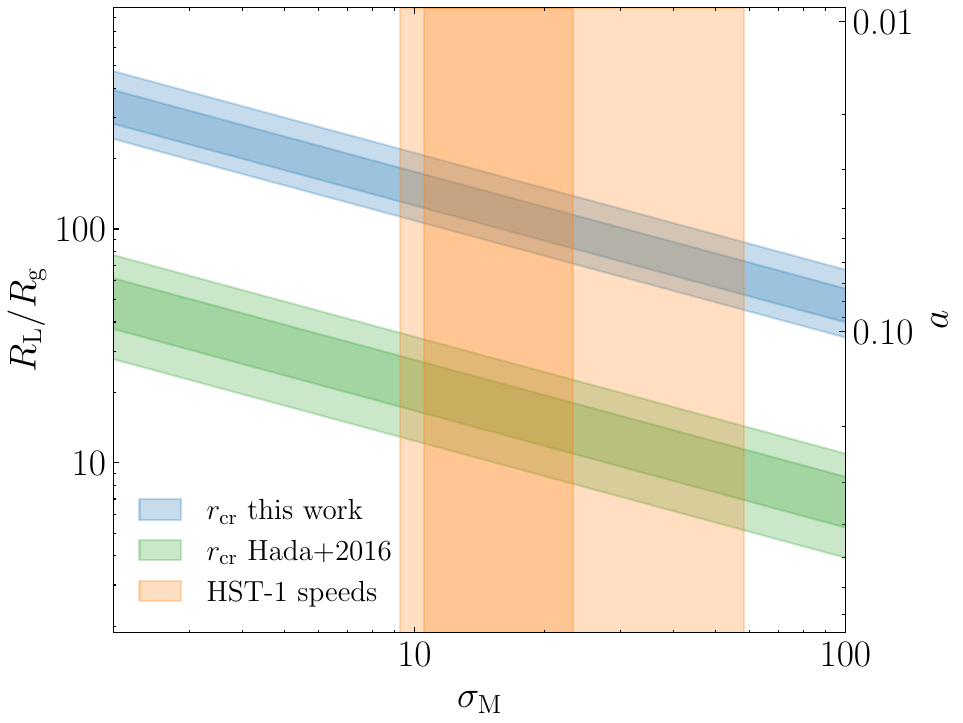}
  \end{center}
  \caption{Constrains on the light cylinder $R_{\rm L}$ (in units of $r_{\rm g}$) and the corresponding dimensionless spin $a$ of M87 SMBH, obtained from the observed radius of the jet at the core appearance position, apparent speeds in HST-1 region and accounting for the uncertainties in the black hole mass, distance and viewing angle estimates. Shade of the color represents 1 and 2 $\sigma$ intervals. The blue color represents the $r_{\rm cr}$ observed in this paper \autoref{sec:searching}. The green color corresponds to the end of the conical expansion profile observed by \citep[][see their fig.9]{2016ApJ...817..131H}.}
\label{fig:constrains_observational}
\end{figure}

\section{Discussion and conclusion}

Thus, it was shown that the formation of a central core near the base of relativistic jets, which was first predicted theoretically and then confirmed by numerical modeling many years ago, should inevitably lead to a change in the dependence of the observed jet width $d(z)$ on the distance $z$ to the central engine. This is due to a sharp decrease in the poloidal magnetic field $B_{\rm p}$ with distance $r$ from the jet axis at large enough distances $z > z_{\rm cr}$ and, as a consequence, with an rapid decrease in the density of radiating particles  $n_{\rm e} \propto B_{\rm p}$ (\ref{ngamma}). As a result, the observed width of the jet $d_{\rm jet}(z)$, which at small distances $z < z_{\rm cr}$  can be estimated as $d_{\rm jet} \approx 2\,r_{\rm jet}$, inevitably becomes noticeably smaller than the geometric size of the jet at larger distances.

Thus, the theory predicts another important property of relativistic jets, which in principle makes it possible to carry out additional diagnostics of their physical parameters. The fact that this issue has not yet been discussed in detail up to now is due to the obvious difficulty of studying the internal structure of the jet at short distances $z \sim z_{\rm cr}$ from the central engine. Note that despite the different literal expressions for conical (\ref{z1con}) and parabolic (\ref{z1par}) flows, the numerical values of $z_{\rm cr}$ for reasonable parameters ($\sigma_{\rm M} \sim 10$--$50$,  $\gamma_{\rm in} \approx 1$, \mbox{$\theta_{\rm jet} \sim 10^{\circ}$)} differ little from each other. As a result, the internal break should be located at a distance of the order of 10--100 $R_{\rm L}$, i.e. only 100--1000 gravitational radii. 

On the other hand, as was already stressed, some indications of the existence of an internal break have already been noted for the closest source M87, for which the observed position of the break corresponds to a fairly large  angular distance $\sim $ 1 mas. For example, Figure 15 given by~\citet{Mertens} shows the break in the dependence of the hydrodynamic Lorentz-factors on the distance $z$ to the central machine. Since in the region of magnetically dominated flow we are considering, the well-known connection $\Gamma = r/R_{\rm L}$ must be satisfied~\citep{MHD}, it can also be interpreted as a break in the dependence of the width of the jet $d_{\rm jet}(z)$. Accordingly, the break in the dependence of the observed jet width $\theta_{\rm jet}$ on distance $z$ is clearly visible in Figure 9 presented by~\citet{2016ApJ...817..131H} at $z_{\rm obs} = 0.6$ mas. 

We also searched for the predicted break in the jet geometry further downstream the jet using stacked image of 37 single-epoch 15 GHz VLBA observations of M87 radio jet from the MOJAVE archive. The transition from almost conical shape at $z_{\rm obs} \approx 6$ mas was found. However, our tests with a synthetic data revealed the presence of the imaging systematical effects that could artificially steepen the expansion profile near the jet origin.

Although the presence of instabilities is generally expected in axisymmetric MHD jets, and the observations of M87 are consistent with development of Kelvin-Helmholtz instability modes~\citep{Lobanov03, 2023MNRAS.526.5949N}, they seem not to disrupt the global jet structure. \citet{Porth15} have explored the stability of a central core, which we connect to the existence of the inner break. It turned out that we should expect the core stability for the ambient pressure $P_\mathrm{ext}\propto z^b$ with $b>2$, which is consistent with our results for the expected from the observations ambient medium pressure power in M87 jet (see Table 2 in~\citealt{NGBNAH19}).

We constrained the spin of the M87 black hole assuming the observed break in the jet shape is due to the appearance of the core of the longitudinal magnetic field. Employing values of the terminal Lorentz-factor inferred from the HST-1 speeds measurements and known viewing angle estimate we constrain the dimensionless spin to be in range 0.05 - 0.1 if the break is located at the $z_{\rm obs} \approx 6$ mas and 0.3 - 0.7 if the break is located upstream at $z_{\rm obs} = 0.6$ mas. The spin constrains dependent on the exact value of the jet magnetization $\sigma_{\rm M}$ among those consistent with HST-1 speeds with higher $\sigma_{\rm M}$ favouring larger spins. 

The estimated range for a black hole spin presented in Figure~\ref{fig:constrains_observational} is in excellent agreement with the previous results, based on the outer break~\citep{NGBNAH19, NKP20}. The corresponding light cylinder radius
coincides with the estimate by~\citet{Kino22}, which was obtained by fitting the observed M87 kinematics. Thus, there are three different methods to constrain the value of $R_\mathrm{L}$: based on kinematics~\citep{Kino22}, fitting the outer~\citep{NGBNAH19} and inner break (this work). All three methods point to the same values of the order of 0.02--0.04 pc (50--100 $R_\mathrm{g}$), which, if the field lines are connected to the black hole ($\Omega_\mathrm{F}=\Omega_\mathrm{H}/2$ holds), corresponds to its rather small non-dimensional spin.

We should also note that the observed inner break in M87 jet requires the ambient medium, collimating the jet, to have a single power-law pressure dependence. This dependence must hold down to the scales of the order of 100-1000 gravitational radii.

\section*{Data availability}
The data underlying this work will be shared on reasonable request to the corresponding author.

\section*{Acknowledgements}
This work was partially supported by the government of the Russian Federation 
(agreement No. 05.Y09.21.0018). 
The analysis of the observational data to estimate the inner break position and the transverse asymmetry analysis are supported by the Russian Science
Foundation: project\footnote{Information about the project: \url{https://rscf.ru/en/project/20-72-10078/}} 20-72-10078.

This research has made use of data from the MOJAVE database that is maintained by the MOJAVE team \citep{2019ApJ...874...43L}. The MOJAVE program is supported under NASA-Fermi grant 80NSSC19K1579.
The National Radio Astronomy Observatory is a facility of the National Science Foundation operated under cooperative agreement by Associated Universities, Inc.

This research made use of \textit{Astropy}, a community-developed core Python package for Astronomy \citep{2013A&A...558A..33A}, \textit{Numpy} \citep{numpy}, \textit{Scipy} \citep{scipy}. \textit{Matplotlib} Python package \citep{Hunter:2007} and \textit{SciencePlots} style \citep{SciencePlots} were used for generating plots in this paper.

\bibliographystyle{mnras}
\bibliography{M87-break}
\label{lastpage}

\end{document}